\documentclass{basi}
\usepackage{epsfig}

\begin{document}
\title[ELODIE classification using ANN]{Automated Classification of ELODIE Stellar Spectral Library Using Probabilistic Artificial Neural Networks}
\author[Bazarghan Mahdi]%
       {Bazarghan Mahdi\thanks{e-mail:mahdi@iucaa.ernet.in} \\
        Inter University Center for Astronomy \& Astrophysics, Post Bag 4, Ganeshkhind, Pune-411007, India}
\maketitle
\label{firstpage}
\begin{abstract}
A Probabilistic Neural Network model has been used for automated
classification of ELODIE stellar spectral library consisting of
about 2000 spectra into 158 known spectro-luminosity classes. The
full spectra with 561 flux bins and a PCA reduced set of 57, 26 and
16 components have been used for the training and test sessions. The
results shows a spectral type classification accuracy of 3.2
sub-spectral type and luminosity class accuracy of 2.7 for the full
spectra and an accuracy of 3.1 and 2.6 respectively with the PCA
set. This technique will be useful for future upcoming large data
bases and their rapid classification.
\end{abstract}

\section{Introduction}

In recent years, Artificial Neural Networks (ANNs) have become very
useful tool for classification type applications (in particular the
stellar spectral classification) in astronomy. Some of the
pioneering efforts of application of ANNs to stellar
spectral classifications are:\\
Gulati et al. (1994a, 1994b), Gulati et al. (1995), Von Hippel et
al. (1994), Weaver et al. (1995),
Singh et al. (1998, 2003), Gupta et al. (2004),\\
and several other groups. All these applications have used ANNs in
various forms based on supervised methods where, the computer is
trained with certain known classes of spectra and then at the
post-training stage, the test set of spectra are input to the
program for a quick classification. The use of Principal Component
Analysis (PCA) greatly reduces the complexity of the training matrix
to a manageable dimension. Most of the previous attempts for
automatic classification of stellar spectra were done using
backpropagation algorithm but in this work probabilistic neural
network Specht D. F., (1990) is presented, which offers some
advantages over the backpropagation technique like: fast training
process, no local minima issues, guaranteed to converge to an
optimal classifier as the size of representative training set
increases, though it suffers from some disadvantages like large
memory and more representative training set requirements.

The section $\S 2$ discusses the input data format; section $\S 3$
outlines the PNN technique; section $\S 4$ discusses the PCA method;
section $\S 5$ gives the performance and results and finally the
section $\S 6$ gives the conclusions of this study.
\section{THE INPUT DATA}

The Jacoby et al. (1984) (hereafter referred as JHC library) was
used as sample spectra for training of the neural network. This
library covers the wavelength range 3510-7427 \AA~ for various O to
M type star which contains 161 spectra of individual stars and 158
spectra were selected from this library as a set of learning
patterns to neural networks. The ELODIE.3 stellar library, 
Prugniel \& Soubiran (2001, 2004) was used as a set of testing
patterns to neural network for classification. This library covers
the wavelength range of 4000 to 6800 \AA~ and contains 1962 spectra
of which 1959 spectra are selected for the classification. The
library is given at two resolutions high and medium of which the
medium resolution is selected. The spectral resolution of JHC is 4.5
\AA~ with one flux value per 1.4 \AA~ and the ELODIE has a
resolution of 0.55 \AA~ with each flux value at 0.2 \AA~. Both these
libraries were brought to a common platform of spectral resolution
of 4.5 \AA~ and sampling fluxes every 5 \AA~ steps (a total of 561
bins) and a wavelength range of 4000-6800 \AA~. This was achieved by
using appropriate convolution and spline fitting routines. Both the
libraries were also normalized to unity to take care of difference
in flux calibration process of two libraries.

\section{PROBABILISTIC NEURAL NETWORKS}

The Probabilistic Neural Network was introduced by Specht D. F.,
(1990). This technique can be used for the classification problem
and estimation of class membership Specht D. F., (1994). This
network is a supervised neural network and its development relies on
Parzen windows classifier and is the
direct continuation of the work of Bayes classifiers 
Specht D. F., (1990), Parzen E. (1962). The Parzen windows method is
a non-parametric procedure that synthesizes an estimation of a
probability density function (pdf) by superposition of a number of
windows.
The PNN learns to approximate the pdf of the training samples.\\
In first step the distance between input vector and the training
input vectors will be evaluated and then a vector will be produced
from which the similarity and closeness of the input to training
input will be known. In the next stage the summation of these
contributions for each class of inputs will take place in order to
produce a vector of probabilities as its net output, and then the
maximum of these probabilities will be selected
\begin{figure}
\epsfig{file=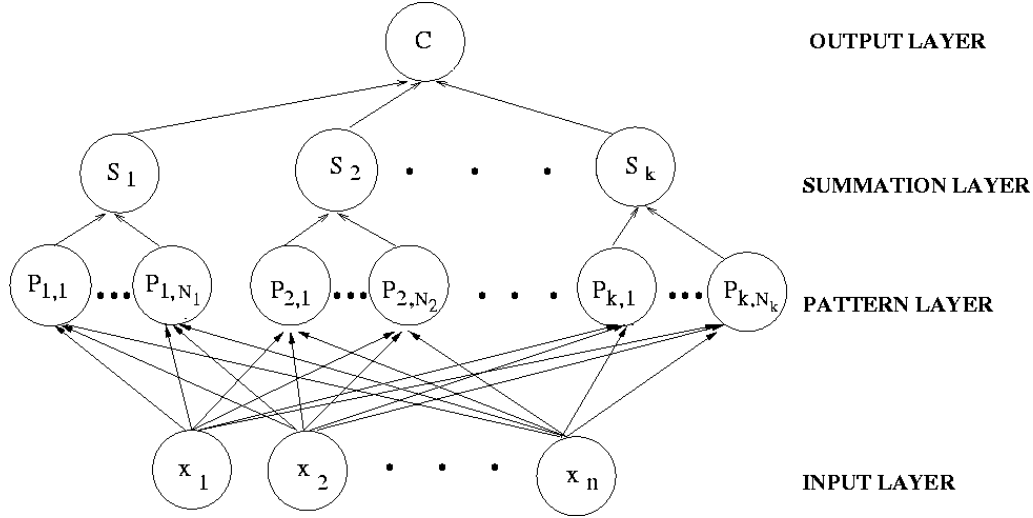,scale=0.55,angle=-90}
\caption{Schematic of a typical Probabilistic Neural Network}
\end{figure}
and will be available at the output.\\
The architecture of a PNN illustrated in Figure 1 with four layers:
the input layer, the pattern layer, the summation layer and the
output layer. An input vector $X = (x_1, x_2, ..., x_n ) \in R^n$ ,
is given to the input neurons and then passed to the pattern layer,
the neurons of which grouped and each group will be dedicated to one
class (or here to one spectra of the training set). The output of
the  $i^{th}$ pattern neuron of $K^{th}$ group in the pattern layer
will be in the form,
\begin{equation}
P_{k,i}(X) = \frac{1}{(2 \pi \sigma^2)^{n/2}} \exp \left( -
\frac{\parallel X - X_{k,i} \parallel^2}{2 \sigma^2} \right),
\end{equation}
where $X_{K,i} \in R^n$   is center of the kernel, $\sigma$ known as
the spread or smoothing parameter. Then the summation layer of the network,
which computes the approximation of the conditional class probability
function given as:
\begin{equation}
S_k (X) = \sum_{i=1}^{N_k} \omega_{k,i} P_{k,i} (X), \hskip 0.5 cm k \in\{ i,....,K\},
\end{equation}
where $N_k$ is the number of neurons of class k, which in our case
$N_k$ will be 561. $\omega_{k i}$ are positive coefficients
satisfying
$\sum_{i=1}^{N_k} \omega_{k i } = 1$.  \\
Then finally at the output layer we will have:
\begin{equation}
C(X) = arg\max_{1 \le k \le K}( S_k ).
\end{equation}

As seen in the Figure 1, the pattern layer is fully connected to the
input layer, with one neuron for each pattern in the training set.
The weight values of the neurons in this layer are set equal to
the different training patterns.  The summation is carried out by
the summation layer neurons. The weights on the connections to the
summation layer are fixed at unity so that the summation layer
simply adds the outputs from the pattern layer neurons. The output
layer neuron produces a binary output value corresponding to the
largest pdf, this indicates the best classification for the
pattern.

\section{PRINCIPAL COMPONENT ANALYSIS}

Principal Component Analysis (PCA) is widely used in signal
processing, statistics and neural computing. The main idea in using
PCA is to extract the components which gives maximum amount of
variance. A $1959\times 561$ matrix, which is input data size with
1959 row as the number of spectra to be classified and 561 is the
size of each spectra. With the help of PCA first tried to reduce the
size of input vectors to simplify the network. The PCA transforms
input vector with highly correlated dimension of 561 variables by
orthogonal transformation to lower number of uncorrelated variables
which are called principal components. The resulting principal
components comes in an ordered manner such that the first principal
component has the largest variation and it will be in reducing order
to the next and subsequent principal components, and it eliminates
the variables with
lowest variation in the whole spectra.\\
This technique is applied to both the training and test data sets
and the best performance obtained with 26 principal components as
shown in the Table 1.
\begin{table}[h]
\begin{center}
\caption{Network performance for PNN with PCA and Full Set}
\vspace{0.3cm}
\begin{tabular}{cccc} \hline

Number & Principal Component & r$^1$ & s$^2$\\ \hline

 1 & 57 & 0.924177 & 452.3178 \\
 2 &  26 & 0.92483 & 448.4866 \\
 3 &  16 & 0.92440 & 450.2248 \\
 4 &  Full spectra (No PCA) & 0.92384 & 454.6563 \\ \hline

 \end{tabular}
 \end{center}
\hspace{2cm} 1  correlation coefficient,  2  standard deviation \\

\end{table}

\section{PERFORMANCE AND RESULT}

\subsection{Spectral class coding}
The performance of the ANN is judged by a quantitative correlation
analysis. The spectro-luminosity classes which are usually in an
alpha-numeric fashion, had to be converted into a numeric code
number (Gulati et al. 1994a) which is unique to each
spectro-luminosity class and is given as follows:

\begin{equation}
Code  Number = 1000.0\times \rm A1  + 100.0\times \rm A2  + ( 1.5 +
2
\times \rm A3 ) \\
\end{equation}
where A1 represents the spectral type of the star (ranging from O to
M which is numbered as 1 to 7); A2 is the sub-spectral type of the
star (with codes ranging from 0.0 to 9.5); and A3 is represents the
luminosity class of the stars (ranging from I to V classes which is
coded into numbers ranging from 0 to 4). For example consider
alpha-numeric spectro-luminosity class F3II which will be coded as
4303.5 and K9.5V will be coded as 6959.5.
\begin{table}[!h]
\begin{center}
\caption{LIST OF SPECTRA APPEARING IN FIGURE 4} \vspace{0.3cm}
\begin{tabular}{ccccc} \hline

Panel & Jacoby & Jacoby & ELODIE & ELODIE \\
 & Spectra & Spectral & Spectra & Spectral \\
  & & Class & & Class \\ \hline

    a  & HD 10032  & 4009.5 & HD 338529 & 2509.5 \\
    b  & SAO 87716 & 3301.5 & HD 190864 & 1705.5 \\
    c  & SAO 87716 & 3301.5 & HD 000108 & 1609.5 \\
    d  & HD 23733  & 3909.5 & HD 172488 & 2059.5 \\
    e  & TR A14    & 5409.5 & HD 017378 & 3501.5 \\
    f  & HD 26514  & 5605.5 & HD 216131 & 7205.5 \\
    g  & HD 12842  & 4301.5 & HD 049330 & 2009.5 \\
    h  & SAO 21536 & 4401.5 & HD 172488 & 2059.5 \\
    i  & BD $58\,^{\circ}$ 0204 & 4201.5 & HD 225160 & 1809.5 \\
    j  & BD $00\,^{\circ}$ 3227 & 4503.5 & HD 172488 & 2059.5 \\
    k  & SAO 21536 & 4401.5 & HD 018409& 1901.5 \\
    l  & SAO 21536 & 4401.5 & HD 192639 & 1809.5 \\
    m  & HD 56030  & 4605.5 & HD 016429& 1955.5 \\
    n  & SAO 21536 & 4401.5 & HD 015558 & 1509.5 \\
    o  & SAO 21536 & 4401.5 & HD 015629 & 1509.5 \\
    p  & HD 107399 & 4909.5 & HD 184499 & 1009.5 \\ \hline

    \end{tabular}

    \end{center}
    \end{table}
\subsection{Performance}
The performance of the classification is evaluated by correlation
analysis. The list of 1959 ELODIE.3 spectra classified by different
networks with different number of principal components was
correlated with respect to the catalog classification given in
ELODIE at the web site:

http://www-obs.univ-lyon1.fr/$\sim$prugniel/soubiran/v3/table\_meas.dat

Table 1 shows the network performance estimated from the linear
correlation coefficient, r, and the standard deviation, s, of the
network and catalog classification. The best performance is for the
network with 26 Principal Components (PCs) with lowest standard
deviation and largest correlation coefficient.
\begin{table}
\caption{LIST OF SPECTRA REJECTED IN PNN WITH PCA}
\vspace{0.3cm}
\begin{center}
\begin{tabular}{c  c  c  c }
\hline

%
Spectra & Catalog & ANN Class & Error \\
Name  & Class &  &  \\
\hline
%
%
%

HD002796 & 4109.5 & 5309.5 & -1200\\
HD014374 & 5009.5 & 6009.5 & -1000\\
HD014626 & 6009.5 & 7009.5 & -1000\\
HD017925 & 6109.5 & 5001.5 &  1108\\
HD019445 & 3409.5 & 4509.5 & -1100\\
HD034078 & 1959.5 & 3301.5 & -1342\\
HD034078 & 1959.5 & 3301.5 & -1342\\
HD038237 & 3309.5 & 4601.5 & -1292\\
HD043823 & 4209.5 & 5409.5 & -1200\\
HD045674 & 4009.5 & 5409.5 & -1400\\
HD048279 & 1809.5 & 2805.5 & -996\\
HD099649 & 5509.5 & 4009.5 & 1500\\
HD099649 & 5509.5 & 4009.5 & 1500\\
HD154543 & 6209.5 & 7305.5 & -1096\\
HD163346 & 3309.5 & 4503.5 & -1194\\
HD184499 & 1009.5 & 4909.5 & -3900\\
HD205811 & 4209.5 & 3109.5 & 1100\\
HD216131 & 7205.5 & 5301.5 & 1904\\
HD218502 & 4309.5 & 2101.5 & 2208\\
HD338529 & 2509.5 & 4009.5 & -1500\\
HD039681 & 2909.5 & 3909.5 & -1000\\
BD+362165 & 5009.5 & 4007.5 & 1002\\
HD000108 & 1609.5 & 2701.5 & -1092\\
HD000108 & 1609.5 & 3301.5 & -1692\\
HD001835 & 5309.5 & 7303.5 & -1994\\
HD008992 & 4601.5 & 5909.5 & -1308\\
HD013267 & 2501.5 & 4201.5 & -1700\\
HD013268 & 1809.5 & 2801.5 & -992\\
HD014947 & 1609.5 & 3909.5 & -2300\\
HD015558 & 1509.5 & 4401.5 & -2892\\
HD015570 & 1409.5 & 4605.5 & -3196\\
HD015629 & 1509.5 & 4401.5 & -2892\\
HD016429 & 1955.5 & 4605.5 & -2650\\
HD016429 & 1955.5 & 4605.5 & -2650\\
HD016429 & 1955.5 & 4605.5 & -2650\\
HD017145 & 2801.5 & 5309.5 & -2508\\
HD017378 & 3501.5 & 5409.5 & -1908\\
HD018409 & 1901.5 & 4401.5 & -2500\\
HD024496 & 5009.5 & 6009.5 & -1000\\
HD034078 & 1959.5 & 3301.5 & -1342\\
\hline
\end{tabular}
\end{center}
\end{table}

\begin{table}
\caption {{\bf{Table 3}}(Contd.)}
\begin{center}
\begin{tabular}{c  c  c  c } \hline

Spectra & Catalog & ANN Class & Error \\
Name  & Class &  &  \\ \hline

HD034078 & 1959.5 & 3301.5 & -1342\\
HD049330 & 2009.5 & 4301.5 & -2292\\
HD053003 & 5001.5 & 6009.5 & -1008\\
HD054908 & 3009.5 & 4009.5 & -1000\\
HD057838 & 6209.5 & 5101.5 & 1108\\
HD089010 & 5207.5 & 7509.5 & -2302\\
HD096094 & 5009.5 & 3805.5 & 1204\\
HD110184 & 5509.5 & 6705.5 & -1196\\
HD157857 & 1709.5 & 2701.5 & -992\\
HD159307 & 4809.5 & 3809.5 & 1000\\
HD161370 & 3009.5 & 4009.5 & -1000\\
HD166734 & 1809.5 & 5001.5 & -3192\\
HD170739 & 2809.5 & 4009.5 & -1200\\
HD172171 & 6105.5 & 7505.5 & -1400\\
HD172488 & 2059.5 & 3909.5 & -1850\\
HD172488 & 2059.5 & 4401.5 & -2342\\
HD172488 & 2059.5 & 4503.5 & -2444\\
HD174512 & 2809.5 & 4503.5 & -1694\\
HD178359 & 4507.5 & 5803.5 & -1296\\
HD182736 & 5009.5 & 6009.5 & -1000\\
HD182736 & 5009.5 & 6009.5 & -1000\\
HD186980 & 1759.5 & 2801.5 & -1042\\
HD190864 & 1705.5 & 3301.5 & -1596\\
HD192639 & 1809.5 & 4401.5 & -2592\\
HD195592 & 1951.5 & 4809.5 & -2858\\
HD199579 & 1609.5 & 2701.5 & -1092\\
HD202124 & 1951.5 & 4301.5 & -2350\\
HD206165 & 2201.5 & 3301.5 & -1100\\
HD210839 & 1609.5 & 3301.5 & -1692\\
HD213470 & 3301.5 & 4309.5 & -1008\\
HD216131 & 7205.5 & 5605.5 &  1600\\
HD216572 & 3009.5 & 4709.5 & -1700\\
HD217086 & 1709.5 & 4605.5 & -2896\\
HD223385 & 3301.5 & 4709.5 & -1408\\
HD225160 & 1809.5 & 3301.5 & -1492\\
HD225160 & 1809.5 & 4201.5 & -2392\\
\hline
\end{tabular}
\end{center}
\end{table}
%


\begin{figure}
\begin{center}
\epsfig{file=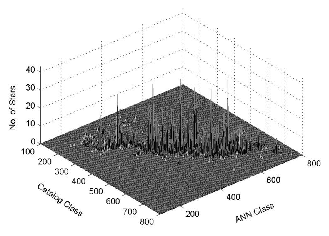,scale=4}
\end{center}
\caption{ELODIE library classification using PNN with PCA before
rejection}
\end{figure}

\begin{figure}
\begin{center}
\epsfig{file=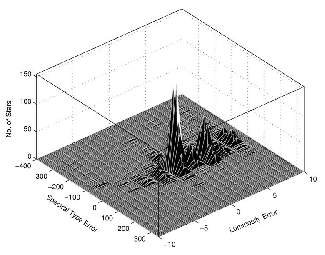,scale=4}
\end{center}
\caption{ELODIE library classification using PNN with PCA before
rejection}
\end{figure}

\begin{figure}

\epsfig{file=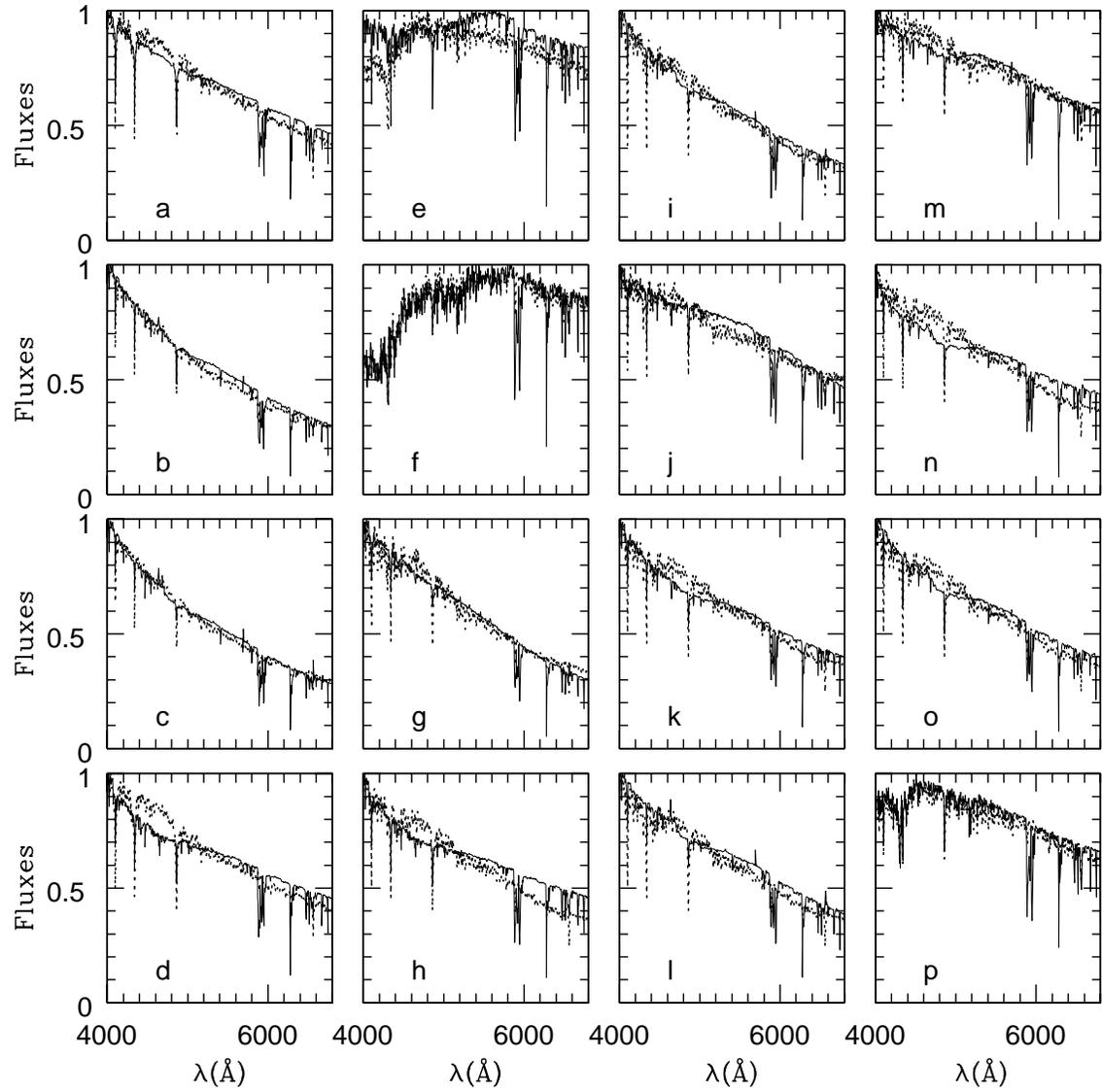,scale=0.8}

\caption{Plots of Jacoby library vs. ELODIE library listed in table
2.}
\end{figure}

\begin{figure}
\begin{center}
\epsfig{file=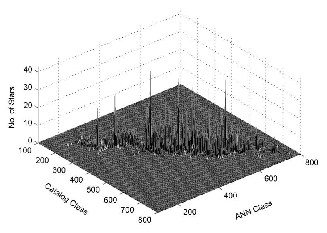,scale=4}
\end{center}
\caption{3-D Scatter diagram of ELODIE library classification using
PNN with PCA after rejection}
\end{figure}

\subsection{Result}
The Neural Network with PNN technique is used to obtain the results
given in Table 1. The result of the classification is in the form,
such that, for each spectra of ELODIE library (test set for ANN),
there is a corresponding spectra from JHC library (training set for
ANN) which is its respective spectro-luminosity class. This result
is given in Table 7 where the name of the star, their spectral and
luminosity class given by ANN and Catalog (the web site given at $\S
5.2$) are presented. The result of the classification of PNN with 26
PCs is presented in a 3D scatter plot format in Figure 2. This
figure also shows the number of spectra available at each spectral
type and luminosity class.

Figure 3 is another 3D picture of classification errors in spectral
type and luminosity class shown on the horizontal axis. Some of the
misclassified spectra with large classification errors are presented
in Figure 4 where, 16 panel shows 16 misclassifications according to
the ANN result. The spectra in dotted lines represent those from the
learning library (JHC) and the spectra in solid lines are the
corresponding ones from the test library (ELODIE).  Table 2
describes the details of Figure 4 with the names of the spectra
plotted in each panel and their coded spectral class. These plots
contribute to the largest classification errors (i.e. high
difference between ELODIE class in Catalog and class obtained by
ANN). But the plots does not support this and shows good match of
these pairs of spectra. So, the classes obtained by ANN could be the
correct class of those spectra.

The result of ANN produced in two schemes:

\indent (1) With PCA:

In this scheme first, PCA reduction technique is applied to reduce
the size of spectra from 561 to 26 PCs and then PNN is used for
automatic classification. As seen from the scatter plot in Figure 2,
there are some spectra which are outliers and contribute to the
large errors in the classification result; these spectra are
rejected and their list is given in Table 3. This table gives the
name of rejected spectra, their class given by catalog and that of
obtained from ANN and the corresponding error.

The accuracy of the classification is evaluated separately for the
Spectral Type (ST) and Luminosity Class (LC) before and after
rejection of the spectra listed in Table 3 from the result of
classification of PNN with PCA, and
the summary of this is given in Table 4.\\

\begin{table}
\caption{ACCURACY OF CLASSIFICATION FOR PNN WITH PCA}
\vspace{0.3cm}
\begin{center}
\begin{tabular}{cc} \hline

Accuracy before rejection & Accuracy after rejection \\ \hline

4.4 ST & 3.1 ST \\
2.7 LC & 2.6 LC \\ \hline

\end{tabular}
\end{center}
\end{table}

After rejecting the spectra listed in Table 3, the 3D scatter plot
is shown in Figure 5.

\indent (2) Without PCA:

In this scheme the PNN is applied to whole spectra of size 561 flux
bins. The list of spectra rejected in this scheme is same as in the
Table 3, in addition with the spectra listed in
Table 5.\\

\begin{table}[!h]
\caption{LIST OF REJECTED SPECTRA FOR PNN WITHOUT PCA}
\vspace{0.3cm}
\begin{center}
\begin{tabular}{cccc} \hline

Spectra & Catalog & ANN Class & Error \\
Name  & Class &  &  \\ \hline

HD016429 & 1955.5 & 4309.5 & -2354\\
HD110184 & 5509.5 & 6603.5 & -1094\\
HD166734 & 1809.5 & 5101.5 & -3292\\
HD172488 & 2059.5 & 4307.5 & -2248\\
HD199579 & 1609.5 & 3001.5 & -1392\\
HD216131 & 7205.5 & 5705.5 &  1500\\ \hline

\end{tabular}
\end{center}
\end{table}

The accuracy of the classification for ST and LC before and after
rejection given in the Table 6.\\

\begin{table}[!h]
\caption{ACCURACY OF CLASSIFICATION FOR PNN WITHOUT PCA}
\vspace{0.3cm}
\begin{center}
\begin{tabular}{c  c } \hline

Accuracy before rejection & Accuracy after rejection \\ \hline

4.5 ST & 3.2 ST \\
2.8 LC & 2.7 LC \\ \hline

\end{tabular}
\end{center}
\end{table}

\section{CONCLUSION}
The classification of ELODIE on the basis of JHC, then comparing the
result of ANN classification with catalog classes and scanning for
the spectra contributing to higher error and plotting them, shows
that though the spectral class given in catalog is different than
that of given by ANN for these plots, but Figure 4 show that they
are of same spectro-luminosity class and it supports ANN
classification result. So by considering these corrections for
misclassification the standard deviation values given in Table 1
will be much lower and also the classification accuracy will
improve. The whole set of ELODIE library is classified with JHC
library as reference to the spectral type accuracy of 3.2 sub
spectral type and luminosity class accuracy of 2.7 for full spectra
and spectral type accuracy of 3.1 sub spectral types and luminosity
class accuracy of 2.6 for PNN with 26 PCs.

The classification of all of the ELODIE test spectra was done in few
seconds, so classification of very large spectral libraries can be
done in considerably short time using this technique in future.

\section*{Acknowledgment}
The author wishes to thank IUCAA for providing the computational
facilities for this work, and Ranjan Gupta for his fruitful
discussion.

\end{document}